\documentclass[aps,prd,superscriptaddress,nofootinbib,eqsecnum,twocolumn]{revtex4}

\pdfoutput=1

\usepackage{amsfonts}
\usepackage{amsmath}
\usepackage{amssymb}
\usepackage{graphicx,color}
\usepackage{float}
\usepackage{hyperref}
\usepackage{subfigure}
\usepackage{mathtools,slashed}
\usepackage{ulem}

\def\be {\begin{equation}}
\def\ee {\end{equation}}
\def\bea {\begin{eqnarray}}
\def\eea {\end{eqnarray}}
\def\bc {\begin{center}}
\def\ec {\end{center}}

\def\nn {\nonumber}

\date{\today}

\begin{document}
\title{Cold QCD at finite isospin density: confronting effective models with recent lattice data}

\author{Sidney S. Avancini}
\affiliation{Departamento de F\'{i}sica, Universidade Federal de Santa Catarina, 88040-900 Florian\'{o}polis, 
Santa Catarina, Brazil}

\author{Aritra Bandyopadhyay}
\affiliation{Departamento de F\'{i}sica, Universidade Federal de Santa Maria, Santa Maria, RS 97105-900, Brazil}

\author{Dyana C. Duarte}
\affiliation{Departamento de F\'{i}sica, Instituto Tecnol\'ogico de 
Aeron\'autica, 12228-900 S\~ao Jos\'e dos Campos, SP, Brazil}

\author{Ricardo L. S. Farias}
\affiliation{Departamento de F\'{i}sica, Universidade Federal de Santa Maria, Santa Maria, RS 97105-900, Brazil}

\begin{abstract}

We compute the QCD equation of state for zero temperature and finite isospin density within the Nambu-Jona-Lasinio model in the mean field approximation, motivated by the recently obtained Lattice QCD 
results for a new class of compact stars: pion stars. 
We have considered both the commonly used Traditional cutoff Regularization 
Scheme and the
Medium Separation Scheme, where in the latter purely vacuum contributions are separated in such a way 
that one is left with 
ultraviolet divergent momentum integrals depending only on  vacuum quantities. We have also compared our 
results with the recent results from Lattice QCD and Chiral Perturbation Theory. 
 
\end{abstract}

\maketitle


\section{Introduction}
\label{sec1}

Quantum chromodynamics (QCD) is the fundamental theory of strong interactions. QCD has a remarkably
rich phase structure with multiple facets which has been vividly explored over the years. Recently,
with the imminent arrival of relativistic Heavy-Ion-Collision (HIC) experiments in FAIR and NICA, 
physical systems at finite baryon densities such as neutron stars have become the ideal subject for 
scrutiny in the heavy ion community~\cite{Fukushima:2010bq,Alford:2007xm}. 
However, systems with finite baryon densities are not easy to deal with theoretically, since in this region 
of QCD phase diagram, first-principle methods such as non-perturbative lattice calculations are not 
accessible due to the well known fermion ``sign problem''~\cite{karsch,Muroya:2003qs}. For a recent
review about the progress of lattice QCD in dealing with sign problem, see Ref~\cite{Bedaque:2017epw}. 

Aside the baryon chemical potential $\mu_B = 3(\mu_u+\mu_d)/2$ (for a 2-flavor system), QCD at finite
density can also be characterized by the isospin chemical potential $\mu_I = (\mu_u-\mu_d)/2$. 
On the contrary to what happens at finite baryonic density, 
systems with finite isospin density does
not suffer from the sign problem and hence are easily accessible to 
lattice QCD based calculations. Initial results of lattice QCD at finite temperature and isospin density
appeared 
in early 2000's~\cite{Kogut:2002zg,Kogut:2002tm} and they were also investigated by
other available techniques, such as chiral perturbation theory
($\chi$PT)~\cite{Son:2000xc,Son:2000by,Splittorff:2000mm,Loewe:2002tw,Loewe:2005yn,
Fraga:2008be,Cohen:2015soa,Janssen:2015lda,Carignano:2016lxe,Lepori:2019vec}, Hard Thermal Loop perturbation 
theory (HTLPt)~\cite{Andersen:2015eoa}, Nambu-Jona-Lasinio (NJL) 
model~\cite{Frank:2003ve,Toublan:2003tt,Barducci:2004tt,He:2005sp,He:2005nk,He:2006tn,Ebert:2005cs,
Ebert:2005wr,Sun:2007fc,Andersen:2007qv,Abuki:2008wm,Mu:2010zz,Xia:2013caa,Khunjua:2018jmn,
Khunjua:2018sro,Khunjua:2017khh,Ebert:2016hkd} and its Polyakov loop extended version 
PNJL~\cite{Mukherjee:2006hq,Bhattacharyya:2012up}, quark meson model 
(QMM)~\cite{Kamikado:2012bt,Ueda:2013sia,Stiele:2013pma,Adhikari:2018cea} and the results were 
largely in qualitative agreement. However, all of the early lattice QCD calculations have been done 
considering unphysical pion masses and/or an unphysical flavour content. Recently, this issue has
been rectified by using an improved lattice action with staggered fermions at physical quark masses 
and the modified lattice QCD results for finite isospin density are presented
in Refs~\cite{Brandt:2016zdy,Brandt:2017zck,Brandt:2017oyy,Brandt:2018wkp}.

In this work we focus on a new type of compact stars, where the pion condensates are considered to be the dominant constituents of the core under the circumstance of vanishing neutron density. Moreover this scenario is easily accessible through first principle methods unlike the study of compact star interiors with high baryon densities. This novel scenario was first identified as pion stars in Ref~\cite{Carignano:2016lxe} and has recently been proposed through lattice QCD in Ref~\cite{Brandt:2018bwq}.

Though pion stars can be described as 
a subset of boson stars~\cite{Wheeler:1955zz,Kaup:1968zz,Jetzer:1991jr,Colpi:1986ye,Liebling:2012fv}, they are free from hypothetical beyond standard model contributions usually associated with boson stars, such as 
QCD axion. Indeed, it can be proved in the framework of a dense neutrino gas that a Bose-Einstein
condensate of positively charged pions can be formed~\cite{Abuki:2009hx}. Further exploration of 
the pion stars' Equation of State (EoS) revealed about its large mass and radius in
comparison with
neutron stars~\cite{Brandt:2018bwq,Andersen:2018nzq}. Recently studies in the similar line have also 
been done within the chiral perturbation theory~\cite{Adhikari:2019mdk}.

Though there are also possibilities of pion condensation in the early universe driven by high lepton assymetry~\cite{Abuki:2009hx,Schwarz:2009ii,Wygas:2018otj}, in 
the current context we will consider the setting of compact stars with zero temperature. Further, the charged pion condensation requires accumulation of 
isospin charge at zero baryon density and zero strangeness. QCD with $\mu_I \neq 0,~\mu_B=\mu_s=T=0$ can 
be and is being realized well within lattice QCD and this new modified lattice result~\cite{Brandt:2018bwq} 
in turn gives us the perfect platform for the consistency check of the effective models mimicking QCD, such 
as NJL model. As emphasized earlier, QCD with finite isospin chemical potential have already been explored
through NJL model, albeit none in light of the new improved lattice results. Additionally 
the present study tries to rectify the regularization issues within NJL model to deal with 
the ultraviolet (UV) divergent momentum integrals. In the Traditional Regularization Scheme (TRS),
commonly used in literature, the sharp UV cutoff $\Lambda$ usually cuts important degrees of freedom
near the Fermi surface leading to incorrect results, specially in scales of the order 
of $\Lambda$, e.g. $\mu_I \sim \Lambda$~\cite{Farias:2005cr,Braguta:2016aov}. On the other 
hand, the Medium Separation Scheme (MSS), coined in Refs~\cite{Farias:2016let,Duarte:2018kfd}, is 
based on a proper separation of medium effects from divergent integrals, originally having 
explicit medium dependence. This results in the disposal of all divergent integrals into the 
pure vacuum part i.e. $\mu_I = 0$ in the current context, as it should be. This scheme has already 
been successfully applied in the context of color superconductivity~\cite{Farias:2005cr} and for
quark matter with a chiral imbalance~\cite{Farias:2016let}. For a proper characterization of 
compact pion stars with high values of $\mu_I$ ($\sim\Lambda$), as we will be dealing with in this work, 
the role of MSS becomes really important in this regard. 

The paper is organized as follows. In section~\ref{sec2} we discuss the basic formalism of
the two-flavor NJL model both within TRS and MSS. In section \ref{sec3} we present our results
obtained with the traditional regularization scheme and with the medium separation scheme, 
thermodynamic results are also presented and contrasted with other state of the art calculations.
We conclude in section \ref{sec4} discussing the aftermath.

\section{Formalism}
\label{sec2}

In this section we revisit the well documented formalism for two-flavor NJL model with finite isospin 
chemical 
potential~\cite{Frank:2003ve,Toublan:2003tt,Barducci:2004tt,He:2005sp,He:2005nk,He:2006tn,Ebert:2005cs,
Ebert:2005wr,Sun:2007fc,Andersen:2007qv,Abuki:2008wm,Mu:2010zz,Xia:2013caa}. We start with the partition
function for the two-flavor NJL model at finite baryonic and isospin chemical potential, given by 
\bea
&&Z_{\textrm{NJL}} (T,\mu_B,\mu_I) = \int[d\bar{\psi}][d\psi]\times \nn\\
&&\exp\left[\int\limits_0^\beta d\tau \int d^3x \left(\mathcal{L}_{\textrm{NJL}} +\bar{\psi} \hat{\mu} \gamma_0 \psi \right)\right], 
\label{partition_f}
\eea
where the quark chemical potential matrix in flavor space is
\bea
\hat{\mu} = \begin{pmatrix}
             \mu_u ~~~~ 0 \\ 0 ~~~~ \mu_d
            \end{pmatrix},
\eea
and $\mu_{u,d}$ can be expressed in terms of the baryonic and the isospin chemical potential as 
\bea
\mu_u &=& \frac{\mu_B}{3} + \mu_I, \nn\\
\mu_d &=& \frac{\mu_B}{3} - \mu_I, \nn
\eea
such that $\mu_B/3 = (\mu_u+\mu_d)/2$ and $\mu_I = (\mu_u-\mu_d)/2$.
$\mathcal{L}_{\textrm{NJL}}$ appearing in Eq.(\ref{partition_f}) is the NJL Lagrangian considering scalar
and pseudoscalar interactions, i.e. 
\bea
\mathcal{L}_{\textrm{NJL}} &=& \bar{\psi}\left(i\slashed{\partial}-m\right)\psi + G \left[\left(\bar{\psi}\psi\right)^2 +\left(\bar{\psi}i\gamma_5 \vec{\tau}\psi\right)^2\right],\nn\\
&=& \bar{\psi}\left(i\slashed{\partial}-m\right)\psi + G \Bigl[\left(\bar{\psi}\psi\right)^2 +\left(\bar{\psi}i\gamma_5 \tau_3\psi\right)^2\nn\\
&&+2\left(\bar{\psi}i\gamma_5 \tau_+\psi\right)\left(\bar{\psi}i\gamma_5 \tau_-\psi\right)\Bigr],
\label{lag_njl}
\eea
where $\psi$ and $m$ represent the quark fields and their current mass respectively and $G$ is the scalar 
coupling constant of the model. $\tau$'s are the generator matrices for the pseudoscalar interactions, which
corresponds to the pionic excitations $\pi_1,\pi_2,\pi_3$ or equivalently $\pi_+,\pi_-,\pi_3$, 
with $\tau_\pm = (\tau_1\pm\tau_2)/\sqrt{2}$.

For finite isospin chemical potential, the isospin symmetry group $SU(2)$ explicitly breaks down to a 
subgroup $U(1)_{I_3}$, third component of the isospin charge $\bf{I}_3$ being the generator~\cite{Mu:2010zz}. So within 
the context of the mean field approximation, for nonzero $\mu_I$ one can consider
the possibility of $\langle\bar{\psi}i\gamma_5 \tau_3\psi\rangle = 0$ as an ansatz, which further 
breaks the $U(1)_{I_3}$ symmetry.
Now we can introduce the chiral condensate $\sigma = -2G\langle\bar{\psi}\psi\rangle$ and pion condensates
\bea
\sqrt{2}\pi_+ &=& -2\sqrt{2}G\langle\bar{\psi}i\gamma_5 \tau_+\psi\rangle =\Delta e^{i\theta},\nn\\
\sqrt{2}\pi_- &=& -2\sqrt{2}G\langle\bar{\psi}i\gamma_5 \tau_-\psi\rangle =\Delta e^{-i\theta},\nn
\eea
where the phase factor $\theta$ indicates the direction of the $U(1)_{I_3}$ symmetry breaking. 
Finally, for the present context of pion stars, we consider $\mu_B=0$, such that $\mu_u=-\mu_d=\mu_I$. Collecting all these information, one can now obtain the thermodynamic potential within the mean field approximation as 
\bea
\Omega_{\textrm{NJL}}(\sigma,\Delta) = \frac{\sigma^2+\Delta^2}{4G}\!-\!2N_c\int_{\Lambda} \frac{d^3k}{(2\pi)^3}\Big[\!E_k^+\!+\!E_k^-\!\Big],
\eea
where $E_k^\pm=\sqrt{\left(E_k\pm\mu_I\right)^2+\Delta^2}$ with $E_k=\sqrt{k^2+M^2}$, $M = m+\sigma$
and the symbol $\int_{\Lambda}$ indicates integrals that need to be regularized. 

The physical values of the condensates vis-a-vis the ground state at finite isospin chemical potential is determined by minimizing $\Omega_{\textrm{NJL}}(\sigma,\Delta)$ with respect to the condensates $\sigma$ and $\Delta$, i.e. by solving the gap equations
\bea
\frac{\partial\Omega_{\textrm{NJL}}}{\partial \sigma}\Big|_{\sigma = \sigma_m} = \frac{\partial\Omega_{\textrm{NJL}}}{\partial \Delta}\Big|_{\Delta = \Delta_m} = 0.
\eea
From these equations we obtain
\bea
\sigma = 4G N_c M \;I_{\sigma}\label{gapS},\\
\Delta = 4G N_c\Delta \;I_{\Delta}\label{gapD},
\eea
with the definitions
\bea
&&I_{\sigma} = \sum_{s= \pm 1}\int_{\Lambda} \frac{d^3k}{(2\pi)^3}
\frac{1}{E_k}\frac{E_k + s\mu_I}{\sqrt{(E_k + s\mu_I)^2 + \Delta^2}},\label{Isig}\\
&&I_{\Delta} = \sum_{s= \pm 1}\int_{\Lambda} \frac{d^3k}{(2\pi)^3}
\frac{1}{\sqrt{(E_k + s\mu_I)^2 + \Delta^2}}.\label{Id}
\eea
In the following subsections we discuss in more details different ways
of regularizing these integrals.
 The thermodynamic quantities, i.e. the pressure, the isospin density and the energy density 
 of the system are then respectively given by
\bea
P_{\textrm{NJL}} &=& - \Omega_{\textrm{NJL}}(\sigma=\sigma_m; \Delta = \Delta_m),\\
\langle n_I\rangle_{\textrm{NJL}} &=& \frac{\partial P_{\textrm{NJL}}}{\partial \mu_I},\\
\varepsilon_{\textrm{NJL}} &=& -P_{\textrm{NJL}} + \mu_I \langle n_I\rangle_{\textrm{NJL}}.
\eea
Finally, the EoS within the two-flavor NJL model is given by the 
relation between $P_{\textrm{NJL}}$ and $\varepsilon_{\textrm{NJL}}$.

\subsection{TRS}

TRS is the most common and used regularization scheme in the literature, as might be seen in 
some good reviews of the NJL model~\cite{reviews}. In this case we just perform the integrations 
in (\ref{Isig}) and (\ref{Id}) up to a cutoff $\Lambda$, that becomes a model parameter. Therefore, 
the gap equations becomes
\bea
\sigma = 4G N_c M\int\limits_0^{\Lambda} \frac{k^2dk}{2\pi^2}
\sum_{j= \pm 1}\frac{E_k + j\mu_I}{E_k\sqrt{(E_k + j\mu_I)^2 + \Delta^2}}\;\;\;\;\\
\Delta = 4G N_c\Delta \int\limits_0^{\Lambda} \frac{k^2dk}{2\pi^2}
\sum_{j= \pm 1}\frac{1}{\sqrt{(E_k + j\mu_I)^2 + \Delta^2}}\;\;\;\;\;\;\;\; 
\eea
This same procedure is used in $\Omega_{\rm NJL}$, that becomes
\be
\Omega_{\textrm{NJL}}^{TRS}(\sigma,\Delta) = \frac{\sigma^2+\Delta^2}{4G}\!-\!2N_c\int\limits_0^{\Lambda} \frac{k^2dk}{2\pi^2}\Big[\!E_k^+\!+\!E_k^-\!\Big]
\ee
and also in the thermodynamic quantities. Specifically, the isospin density becomes
\be
\langle n_I\rangle_{\textrm{NJL}}^{\rm TRS} = -2N_c \int\limits_0^{\Lambda} \frac{k^2dk}{2\pi^2}
\left[\frac{E_k - \mu_I}{E_k^-} - \frac{E_k + \mu_I}{E_k^+}\right]\;.
\ee
\subsection{MSS}

Since NJL is nonrenormalizable, any physical quantity will depend on the scale of the model $\Lambda$. However, it is very important
to keep in mind that cutoff dependent  medium terms due to a naive regularization of the integrals 
may lead to results completely different from the ones obtained with a more careful treatment 
of divergences. MSS provides a tool to disentangle medium dependence from divergent contributions,
so that only vacuum integrals need to be regularized. This scheme has been applied 
to the NJL model and successfully shows qualitative agreement with lattice simulations and 
more elaborated theories, as might be seen in Refs.~\cite{Farias:2005cr,Farias:2016let,Duarte:2018kfd}.

The implementation of MSS starts by rewriting, for example, $I_{\Delta}$ given in Eq.~(\ref{Id}) as
\be
I_{\Delta} = \frac{1}{\pi}\sum_{j= \pm 1}\int\limits_{-\infty}^{+\infty}dx\int_{\Lambda} \frac{d^3k}{(2\pi)^3}
\frac{1}{x^2 + (E_k + j\mu_I)^2 + \Delta^2}\;.\label{Id4}
\ee
Using the identity
\bea
\lefteqn{ \frac{1}{x^2 + (E_k + j\mu_I)^2 + \Delta^2}} \nn\\
&& = \frac{1}{x^2 + k^2 + M_0^2}\nn\\
&& + \frac{M_0^2 - \Delta^2 - \mu_I^2 - M^2 - 2j\mu_I E_k}{(x^2 + k^2 + M_0^2)
\left[x^2 + (E_k + j\mu_I)^2 + \Delta^2\right]}\label{ident}
\eea
(where $M_0$ is the vacuum mass, when $\mu_I = \Delta = 0$) we obtain, after two iterations,
\bea
\lefteqn{\sum_{j= \pm 1}\frac{1}{x^2 + (E_k + j\mu_I)^2 + \Delta^2}}\nn\\
&& = \frac{2}{x^2 + k^2 + M_0^2} + \frac{2\mathcal{M}}{(x^2 + k^2 + M_0^2)^2}\nn\\
&& + \frac{2\mathcal{M}^2 + 8\mu_I^2 E_k^2}{(x^2 + k^2 + M_0^2)^3}\nn\\
&& + \sum_{j= \pm 1}\frac{(\mathcal{M} - 2j\mu_I E_k)^3}{(x^2 + k^2 + M_0^2)^3
\left[x^2 + (E_k + j\mu_I)^2 + \Delta^2\right]}\nn\\
\eea
where we have defined $\mathcal{M} = M_0^2 - \Delta^2 - \mu_I^2 - M^2$. After some manipulations 
and performing the integration in $x$ indicated in (\ref{Id4}) we obtain
\bea
\lefteqn{I_{\Delta}^{\rm MSS} = 2I_{\rm quad} - (M^2 - M_0^2 + \Delta^2 - 2\mu_I^2)I_{\rm log}}\nn\\
&&+ \left[\frac{3(\mathcal{M}^2 + 4\mu_I^2M^2)}{4} - 3\mu_I^2M_0^2\right]I_1 + 2I_2
\eea
with the definitions
\bea
I_{\rm quad} & = & \int\frac{d^3k}{(2\pi)^3}\frac{1}{\sqrt{k^2 + M_0^2}}\;,\\
I_{\rm log} & = & \int\frac{d^3k}{(2\pi)^3}\frac{1}{(k^2 + M_0^2)^{\frac{3}{2}}}\;,\\
I_1 & = & \int\frac{d^3k}{(2\pi)^3}\frac{1}{(k^2 + M_0^2)^{\frac{5}{2}}}\;,\\
I_2 & = & \frac{15}{32}\sum_{j= \pm 1}\int\frac{d^3k}{(2\pi)^3}\int\limits_0^1 dt (1-t)^2\nn\\
&&\times\frac{(\mathcal{M} - 2j\mu_I E_k)^3}{\left[(2j\mu_I E_k - \mathcal{M})t + k^2 + M_0^2\right]^{\frac{7}{2}}}\;,
\label{i_2}
\eea
where, in the last line of  the equation above we have used the Feynman parametrization
\be
\frac{1}{A_1^n A_2^m} = \frac{\Gamma(n+m)}{\Gamma(n)\Gamma(m)}
\int\limits_0^1 dt\frac{t^{n-1}(1-t)^{m-1}}{\left[A_1 t + A_2(1-t)\right]^{n+m}}\;.
\ee
Using similar steps one may write
\bea
\lefteqn{I_{\sigma}^{\rm MSS} = 2I_{\rm quad} - (M^2 - M_0^2 + \Delta^2)I_{\rm log} + I_3}\nn\\
&& + 3\left[\frac{\mathcal{M}^2}{4} + \mu_I^2(M^2 - M_0^2 - \mathcal{M})\right]I_1 + 2I_2,
\eea
with
\bea
I_3 & = & \frac{15}{16}\sum_{j = \pm 1}\int\frac{d^3k}{(2\pi)^3}
\int\limits_0^{\infty}\frac{t^2 dt}{\sqrt{1+t}}\nn\\
&& \times \frac{1}{E_k} \frac{j\mu_I(\mathcal{M} - 2j\mu_I E_k)^3}
{\left[(k^2 + M_0^2)t + (E_k + j\mu_I)^2 + \Delta^2\right]^{\frac{7}{2}}}\;.
\eea
Using MSS the expression for the normalized thermodynamic potential becomes
\bea
\lefteqn{\Omega_{\rm NJL}^{\rm MSS}(\sigma,\Delta) = 
\frac{\sigma^2+\Delta^2}{4G}}\nn\\
&&-\!2N_c\Biggl\{\tilde{\mathcal{M}}I_{\rm quad}
 - \frac{1}{4}\left(\tilde{\mathcal{M}^2} - 4\mu_I^2\Delta^2\right)I_{\rm log}\nn\\
&&+\int\frac{d^3k}{(2\pi)^3}\left[\frac{\tilde{\mathcal{M}^2} - 4\mu_I^2\Delta^2}{4E_{k,0}^3}
- \frac{\tilde{\mathcal{M}}}{E_{k,0}}\right.\nn\\
&&- 2E_{k,0} + E_k^+ + E_k^-\Biggl]\Biggl\}
\eea
with the definitions $\tilde{\mathcal{M}} = \Delta^2 + M^2 - M_0^2$ and $E_{k,0} = \sqrt{k^2 + M_0^2}$.
To obtain the expression for the isospin density we follow the same procedure used
for the calculation of $I_{\Delta}$ 
and $I_{\sigma}$, but due to its different divergency structure 
we need to iterate the identity (\ref{ident}) once more. The final expression is
\bea
\lefteqn{\langle n_I\rangle_{\textrm{NJL}}^{\rm MSS} = 2\mu_I\Delta^2 I_{\rm log}} \nn\\
&&3\mu_I\left[\frac{\mathcal{M}^2}{4} + \mathcal{M}(M_0^2 - M^2) + M^2\mu_I^2 
+ \frac{2\mu_I^2M_0^2}{3} \right]I_1\nn\\
&& + 2\mu_I I_2 - \frac{5\mu_I M^2}{4}\left[3\mathcal{M}^2 + 4\mu_I^2M^2\right]I_4\nn\\
&& + \frac{5\mu_I}{4}\left(4\mu_I^2(M_0^2-2M^2) - 3\mathcal{M}^2\right)I_5 + I_6
\eea
with the remaining definitions,
\bea
I_4 & = & \int\frac{d^3k}{(2\pi)^3}\frac{1}{(k^2 + M_0^2)^{\frac{7}{2}}}\;,\\
I_5 & = & \int\frac{d^3k}{(2\pi)^3}\frac{k^2}{(k^2 + M_0^2)^{\frac{7}{2}}}\;,\\
I_6 & = & \frac{35}{32}\sum_{j = \pm 1}\int\frac{d^3k}{(2\pi)^3}
\int\limits_0^{\infty}\frac{t^3 dt}{\sqrt{1+t}}\nn\\
&& \times \frac{jE_k(\mathcal{M} - 2j\mu_I E_k)^4}
{\left[(k^2 + M_0^2)t + (E_k + j\mu_I)^2 + \Delta^2\right]^{\frac{9}{2}}}\;.
\eea
Note that integrals $I_1$ to $I_6$ are all finite, and must be performed up to infinite in $k$.
This is the fundamental difference between TRS, where we cut the whole integral in the 
cutoff $\Lambda$ and MSS, where all finite medium contributions are separated and performed 
for the whole momentum range.

\section{Results}
\label{sec3}

The parameter set used for the purpose of the present study are 
$m=4.76$ MeV, $\Lambda= 659$ MeV and $G=4.78$ GeV$^{-2}$ which we have obtained by fitting the same value 
of the pion mass as used by Lattice QCD~\cite{Endrodi:pc}, i.e. $m_\pi = 131.7$ MeV, and other 
parameters as $f_\pi = 92.4$ MeV and $\langle \bar{\psi}\psi\rangle^{1/3}=-250$ MeV. This values corresponds
to a vacuum mass $M_0 \simeq 303.5$ MeV.  

\begin{figure}[h!]
 \begin{center}
 \includegraphics[scale=0.32]{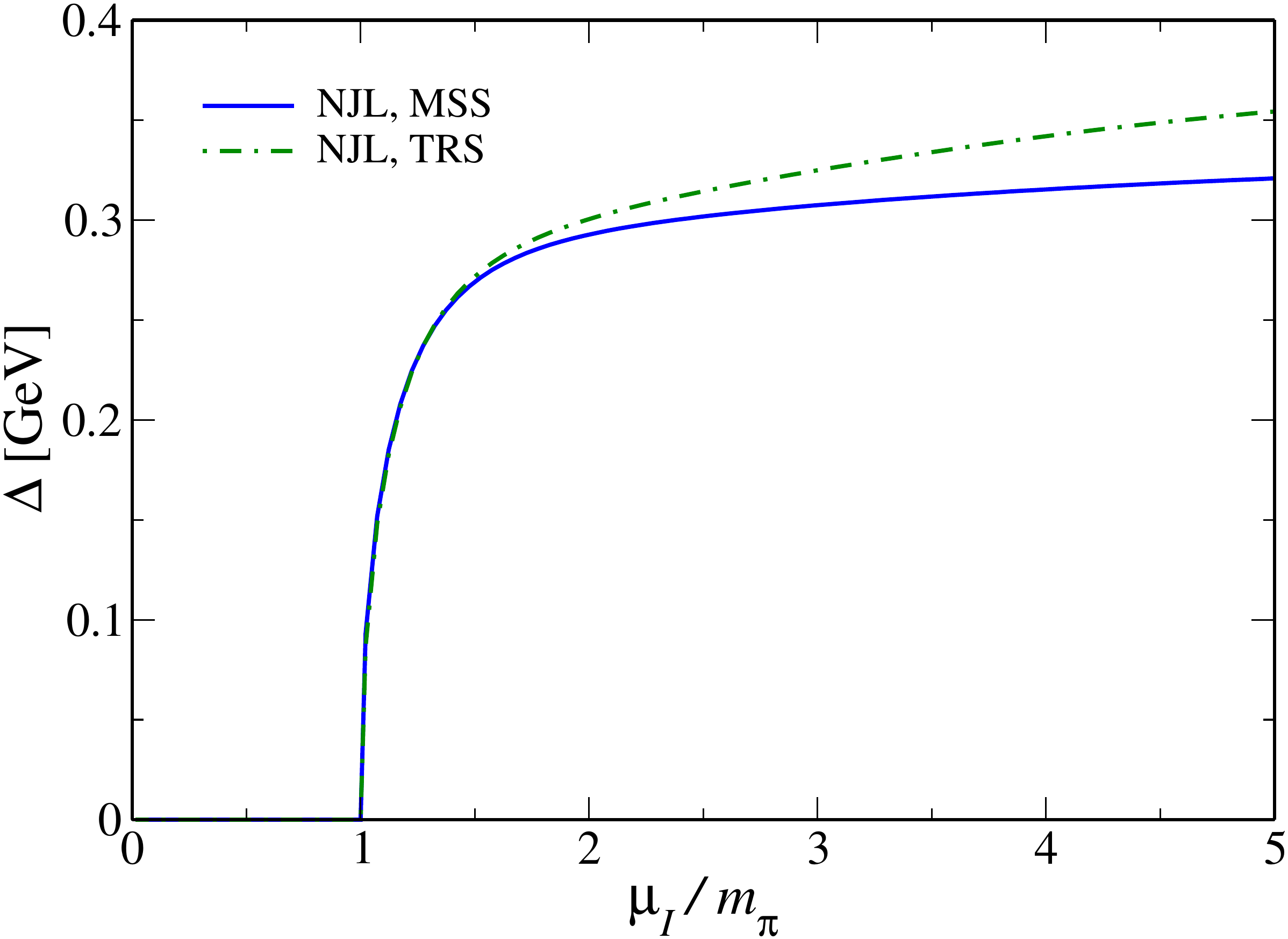}
 \caption{Variation of the amplitude of the pion condensate $\Delta$ as a function of the normalized
 isospin chemical potential $\mu_I/m_\pi$, using both TRS and MSS.}
 \label{vary_Delta}
 \end{center}
\end{figure}
Figure \ref{vary_Delta} shows the variation of the pion condensate $\Delta$ with $\mu_I$, scaled by 
the  pion mass value.  As might be seen from the plot, higher 
values of $\mu_I$ (starting from $\mu_I \sim 1.5 m_\pi$) draw 
the differences between the two regularization processes.
Notice that the values of  $\Delta$ are increasingly larger for TRS than MSS when $\mu_I$ grows.
At $\mu_I\sim\Lambda$ (i.e. $\mu_I\sim 5m_\pi$) the difference between TRS and MSS goes up to $30-35$ MeV. 
This difference in $\Delta$ at higher values of $\mu_I$ also justifies the use of the medium separation
scheme, specially since we are working at the zero temperature limit.

In the following part of this section we shall discuss our results for different relevant thermodynamic
quantities within the two-flavor  NJL model, comparing each one with the
corresponding recent Lattice QCD 
results~\cite{Brandt:2018bwq} and Chiral perturbation theory~\cite{Adhikari:2019mdk} results for both
Leading Order (LO) and Next to Leading Order (NLO). It is important to mention that in the present study we are using data sets collected through private communications~\cite{Adhikari:pc}. In the $\chi$PT results used in this study the authors have used the Particle Data Group (PDG) value of the $f_\pi$, i.e. $\sqrt{2}f_\pi = 130.2 ~(\pm 1.7)$ MeV and for the pion mass $m_{\pi}=135$ MeV. Due to the uncertainty in the values of the low-energy constants~\cite{Adhikari:2019mdk,Adhikari:pc} the uncertainty for the $\chi$PT-NLO results have also been presented.
\begin{figure*}[!]
 \begin{center}
 \subfigure{\includegraphics[scale=0.32]{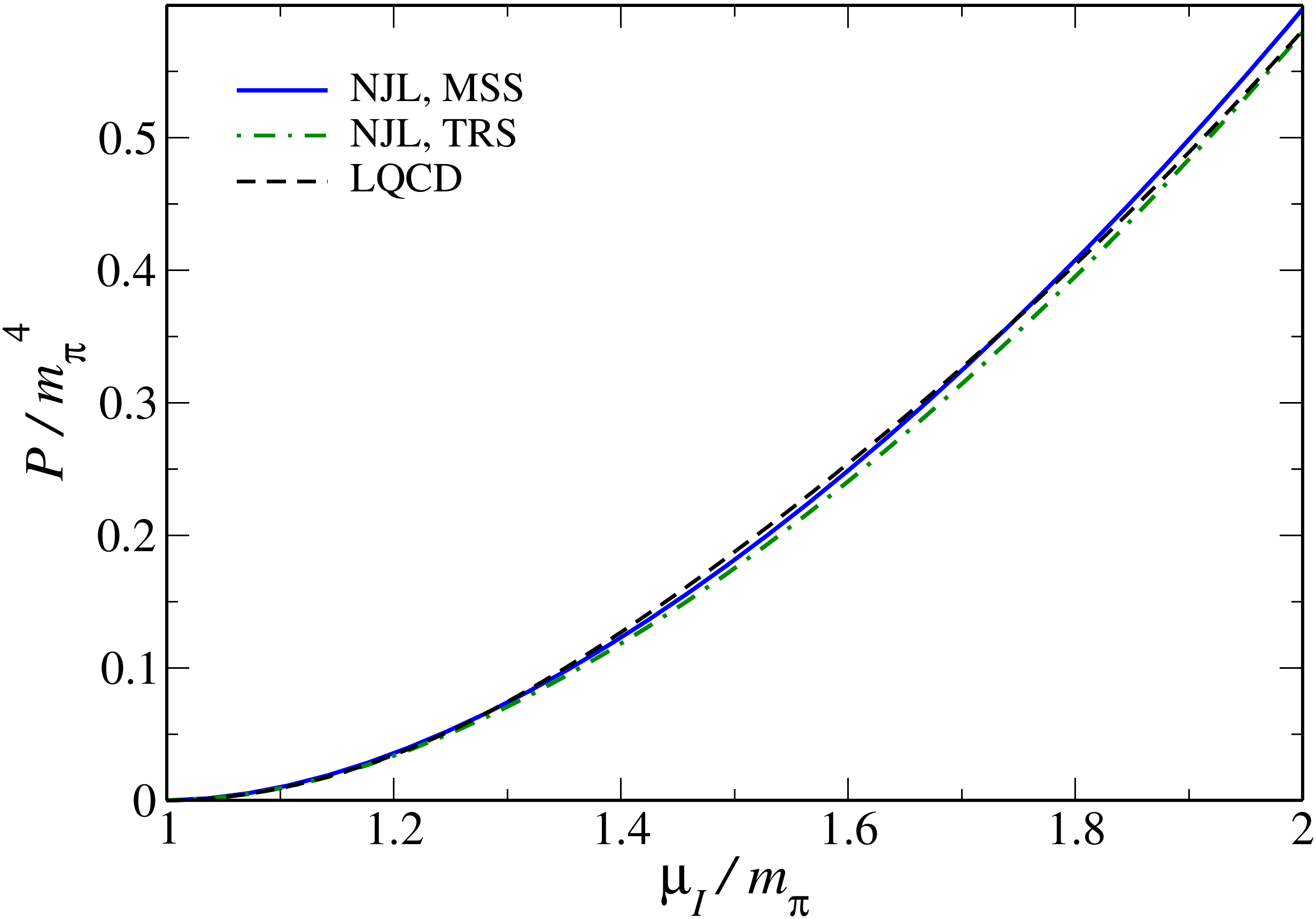} }\hspace{0.5cm}
 \subfigure{\includegraphics[scale=0.32]{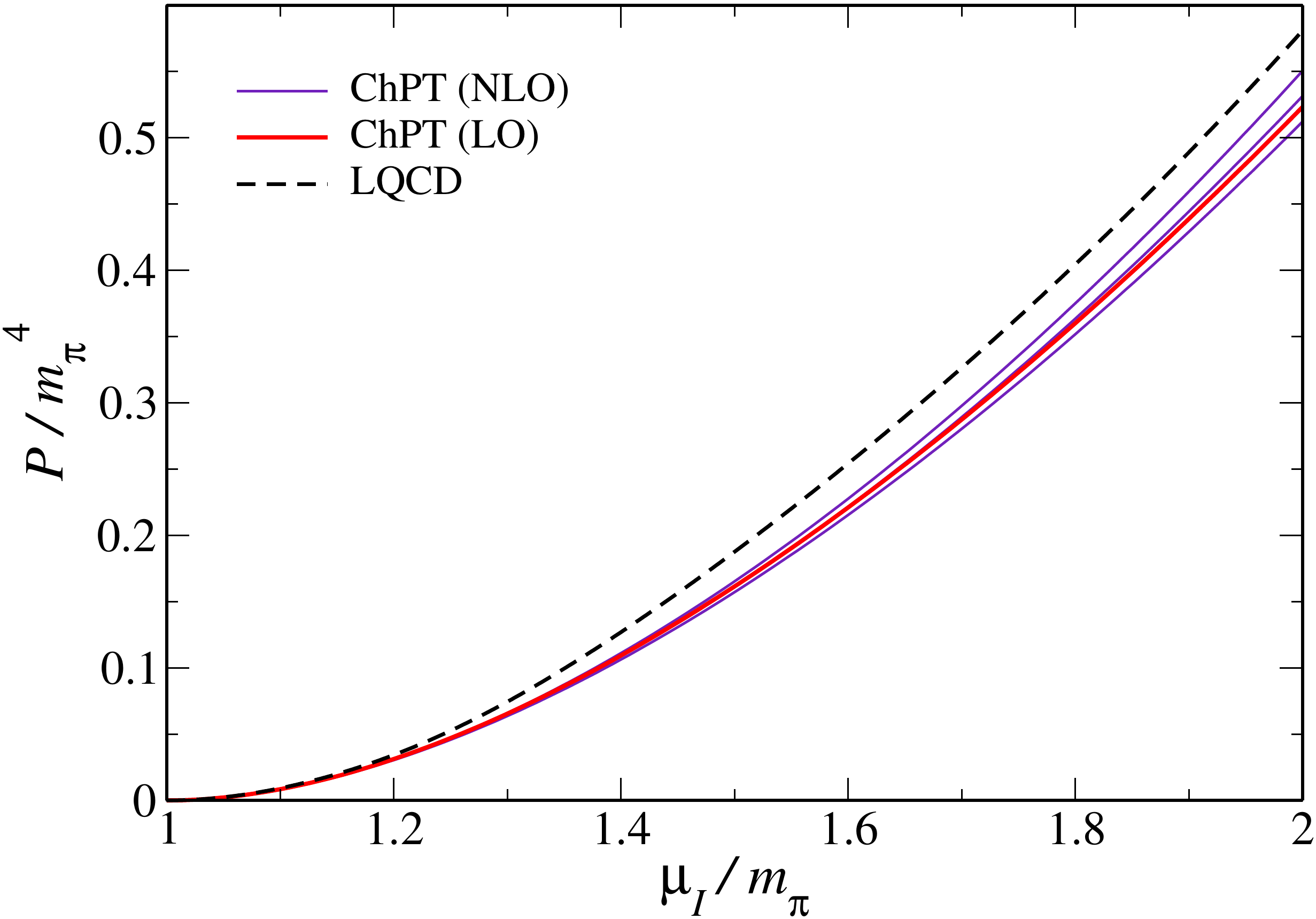} }
 \caption{Variations of the normalized pressure ($P/m_\pi^4$) as a function of the normalized isospin chemical potential $\mu_I/m_\pi$. The LQCD results~\cite{Brandt:2018bwq} have been compared with the behavior of MSS and TRS within the NJL model (left panel) and with up to NLO results within $\chi$PT~\cite{Adhikari:pc} (right panel). Both the plots are specifically zoomed into the region of interest, up to the value of $\mu_I$ for which LQCD data is available. The three lines for $\chi$PT-NLO depicts the uncertainty in the result due to the uncertainty in the low-energy constants~\cite{Adhikari:2019mdk,Adhikari:pc}.}
 \label{pressure_comparison}
 \end{center}
\end{figure*}
\begin{figure*}[!]
 \begin{center}
 \subfigure{\includegraphics[scale=0.32]{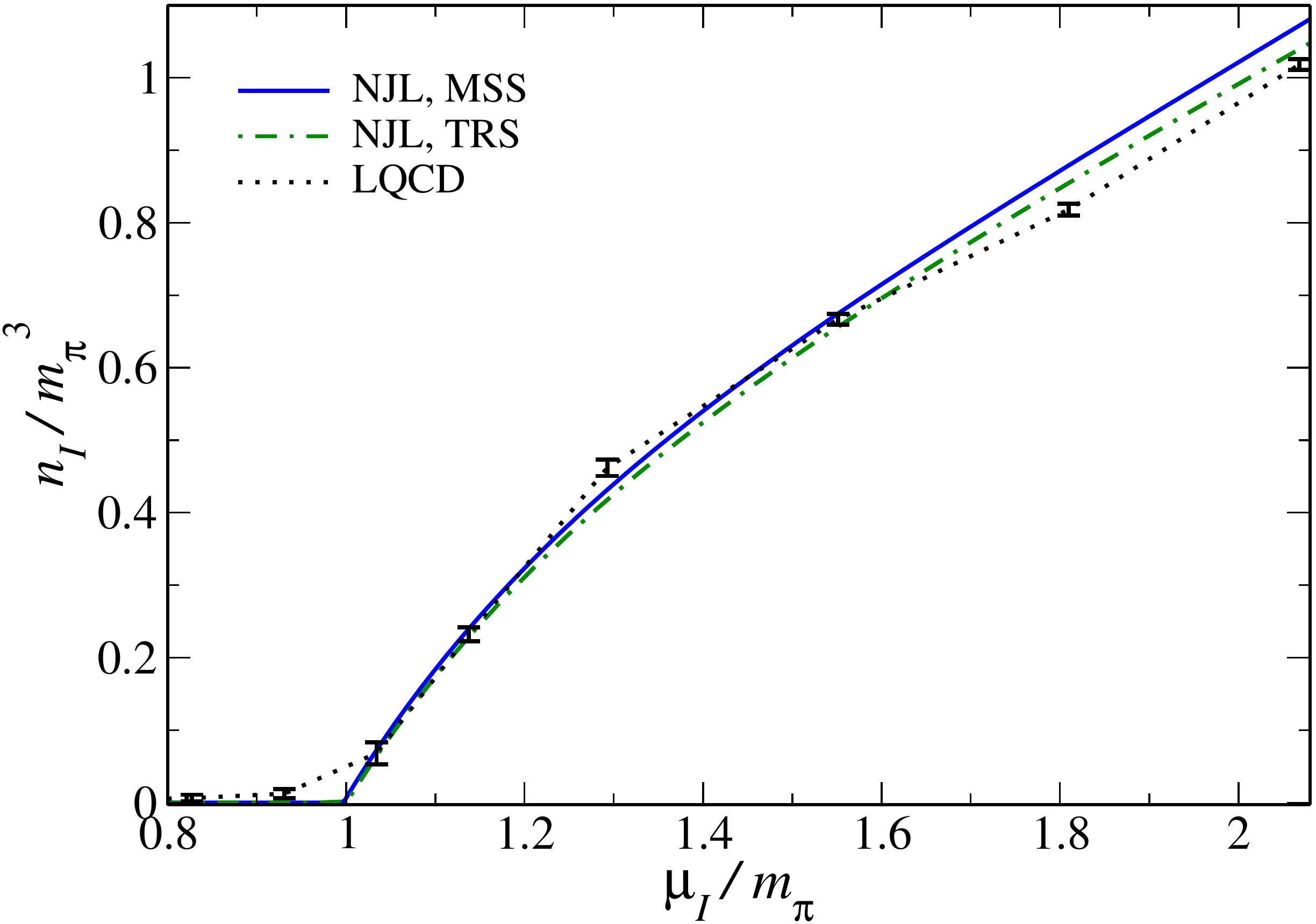}}\hspace{0.5cm}
 \subfigure{\includegraphics[scale=0.32]{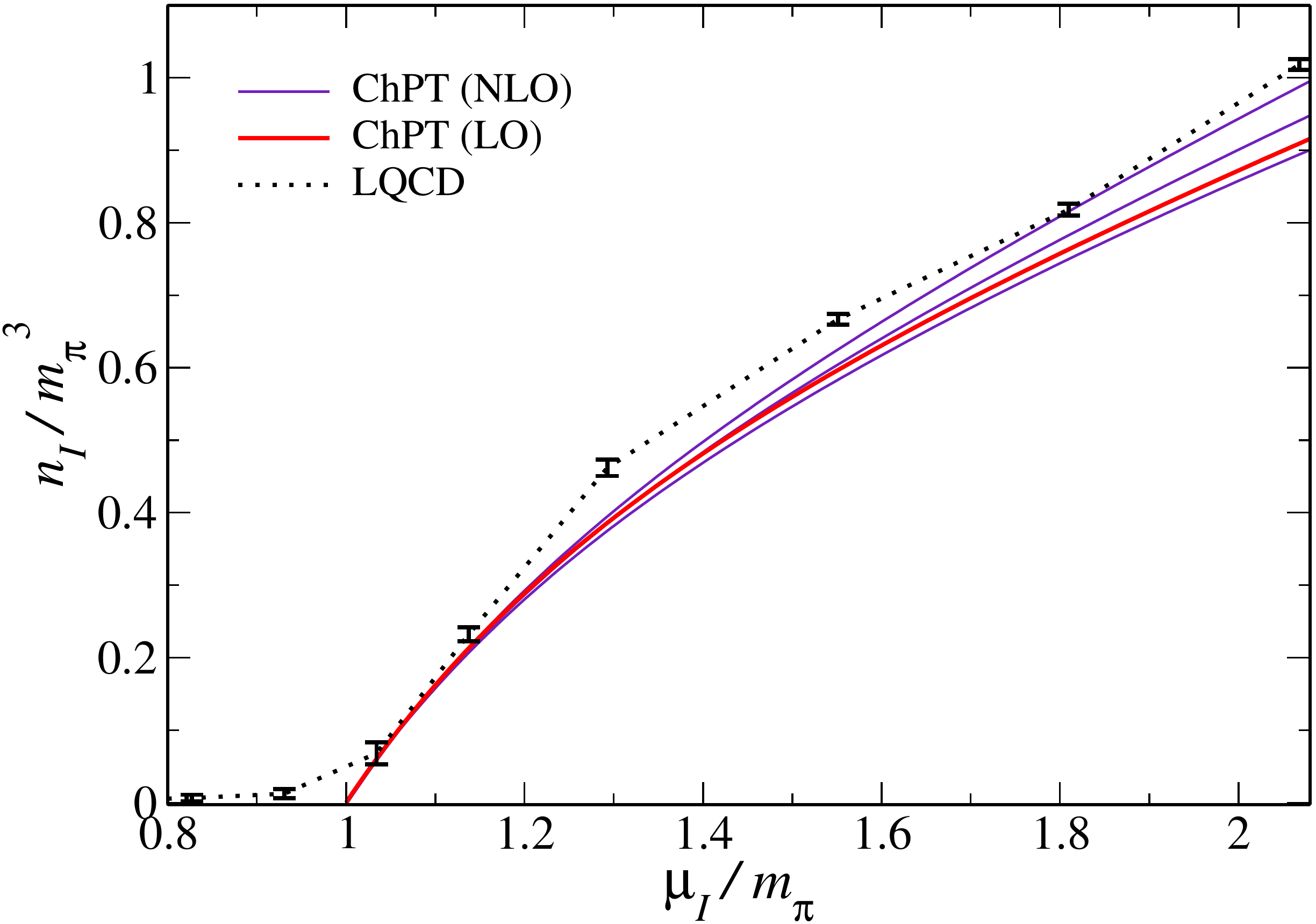}}
  \caption{Variations of the normalized isospin density ($n_I/m_\pi^3$) as a function of the normalized isospin chemical potential $\mu_I/m_\pi$. The LQCD results~\cite{Brandt:2018bwq} have been compared with the behavior of MSS and TRS within the NJL model (left panel) and with up to NLO results within $\chi$PT~\cite{Adhikari:pc} (right panel). The plots are specifically zoomed into the region of interest, up to the value of $\mu_I$ for which LQCD data is available. The three lines for $\chi$PT-NLO depicts the uncertainty in the result due to the uncertainty in the low-energy constants~\cite{Adhikari:2019mdk,Adhikari:pc}. Unlike the other thermodynamic quantities, here relatively fewer amount of lattice data points are shown with respective error bars. The dotted line represents the first order interpolation of the latter.}
  \label{ni_comparison}
 \end{center}
\end{figure*}
\begin{figure*}[!]
 \begin{center}
  \subfigure{\includegraphics[scale=0.32]{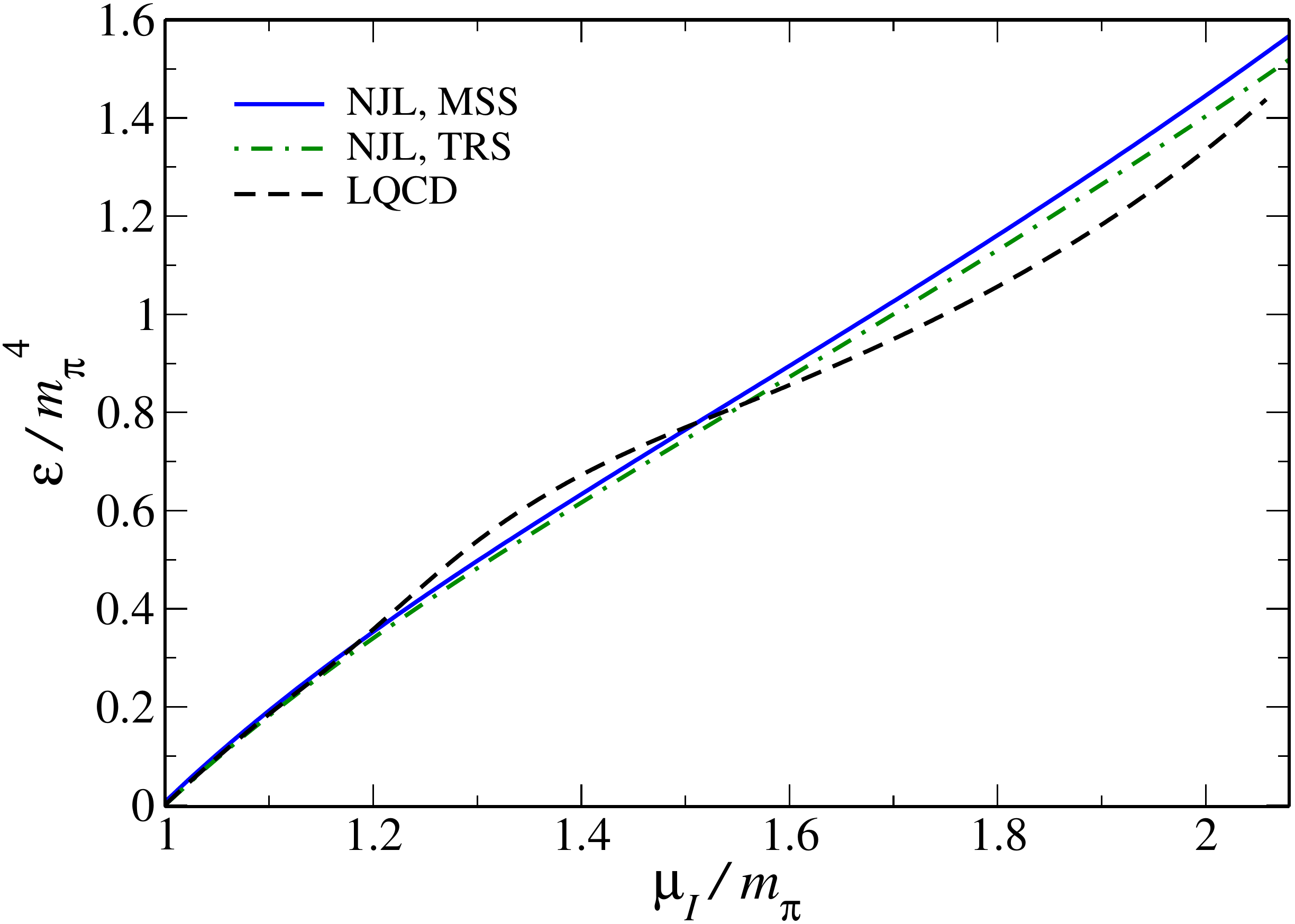} }\hspace{0.5cm}
  \subfigure{\includegraphics[scale=0.32]{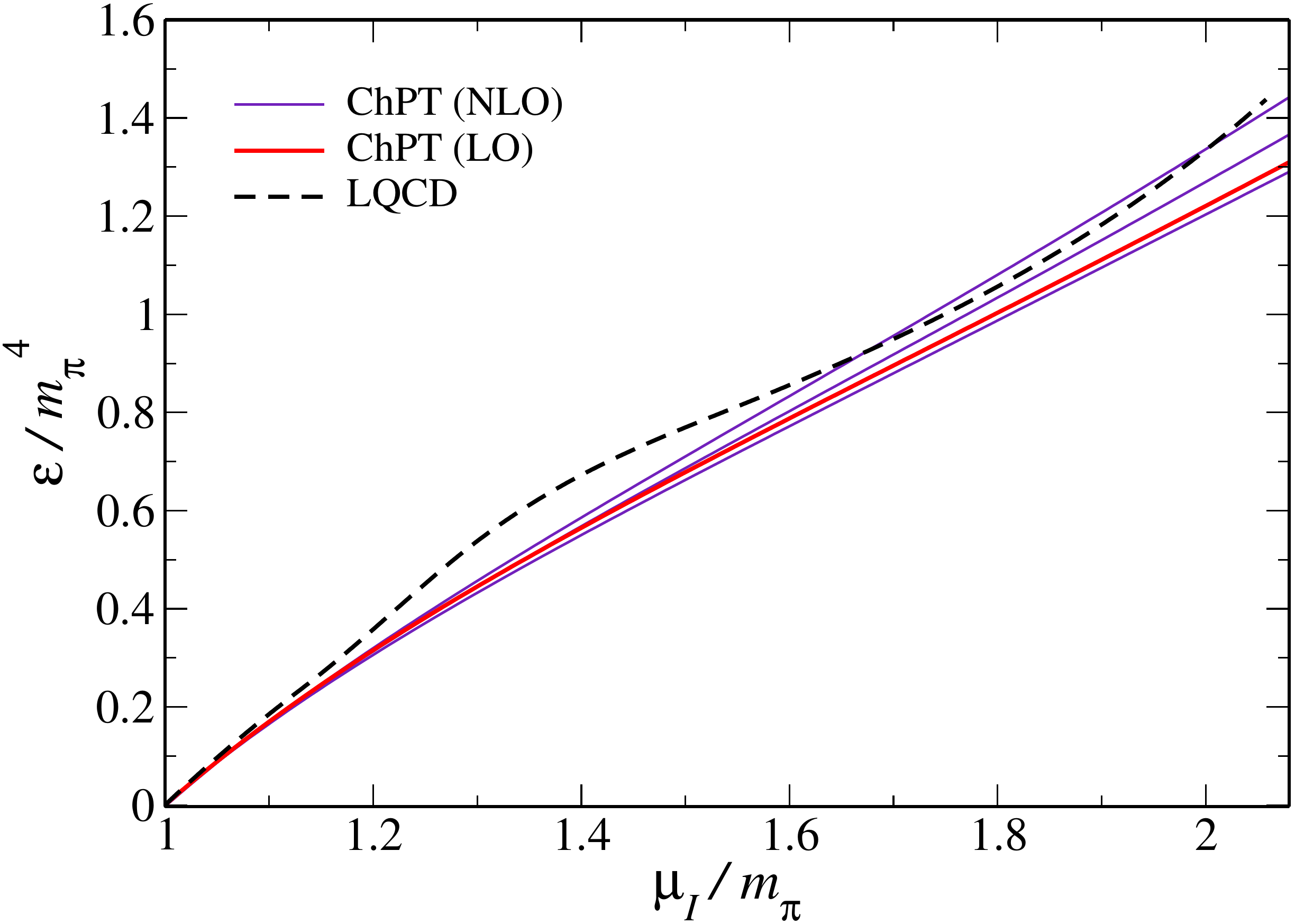} }
   \caption{Variations of the normalized energy density ($\varepsilon/m_\pi^4$) as a function of the normalized isospin chemical potential $\mu_I/m_\pi$. The LQCD results~\cite{Brandt:2018bwq} have been compared with the behavior of MSS and TRS within the NJL model (left panel) and with up to NLO results within $\chi$PT~\cite{Adhikari:pc} (right panel). The plots are specifically zoomed into the region of interest, up to the value of $\mu_I$ for which LQCD data is available. The three lines for $\chi$PT-NLO depicts the uncertainty in the result due to the uncertainty in the low-energy constants~\cite{Adhikari:2019mdk,Adhikari:pc}.}
  \label{ed_comparison}
 \end{center}
\end{figure*}
In figures \ref{pressure_comparison}, \ref{ni_comparison} and \ref{ed_comparison}, respectively,
the variations of normalized pressure, isospin density and energy density are shown with respect to the isospin
chemical potential scaled by $m_\pi$. These plots have mainly focussed on the region where 
$m_\pi \lesssim\mu_I\lesssim2m_\pi$ as the region of interest, throughout which lattice QCD data was 
available\footnote{In general within Lattice QCD calculations, the maximum value of $\mu_I$ is constrained
by the value of the lattice spacing.}. In this range of $\mu_I$, the difference in results for TRS and MSS is
relatively small, as evident from the plots. Comparing NJL results we can observe that TRS has an infinitesimally better agreement with current LQCD than MSS.  LO and NLO results within $\chi$PT have also been
compared among others. Figure \ref{pressure_comparison} distinctively shows the comparability between
the NJL and LQCD results, specially in comparison with $\chi$PT results up to NLO. Note that for the $\chi$PT datasets used here, the value of pion mass used was taken as $135$ MeV (particle data group). Using instead a pion mass closer to the value adopted by LQCD, i.e, $m_\pi = 131 \pm 3$ MeV and 
$\sqrt{2}f_\pi = 128 \pm 3$ MeV, as it is made in the published version of Ref.~\cite{Adhikari:2019mdk}, the agreement between LQCD and $\chi$PT is improved.
Figures \ref{ni_comparison} and \ref{ed_comparison} show a typical behavior of LQCD data, which cross
over the NJL TRS and MSS results around $\mu_I \sim 1.5 m_\pi$, though overall being largely in agreement. 
This cross over could be due to the current unavailability of larger number of lattice data for isospin density. 

\begin{figure*}[!]
 \begin{center}
  \subfigure{\includegraphics[scale=0.33]{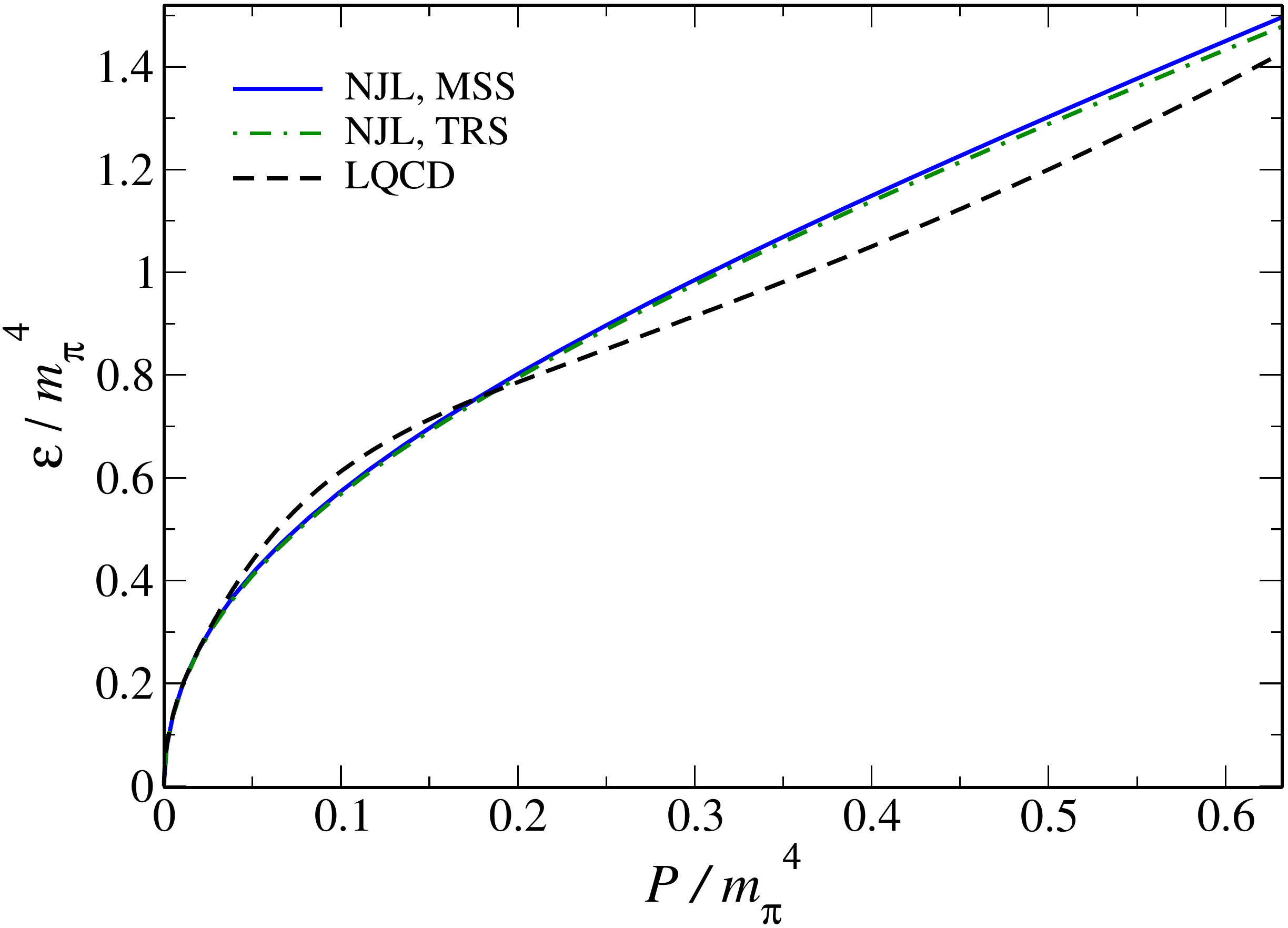}}\hspace{0.5cm}
  \subfigure{\includegraphics[scale=0.33]{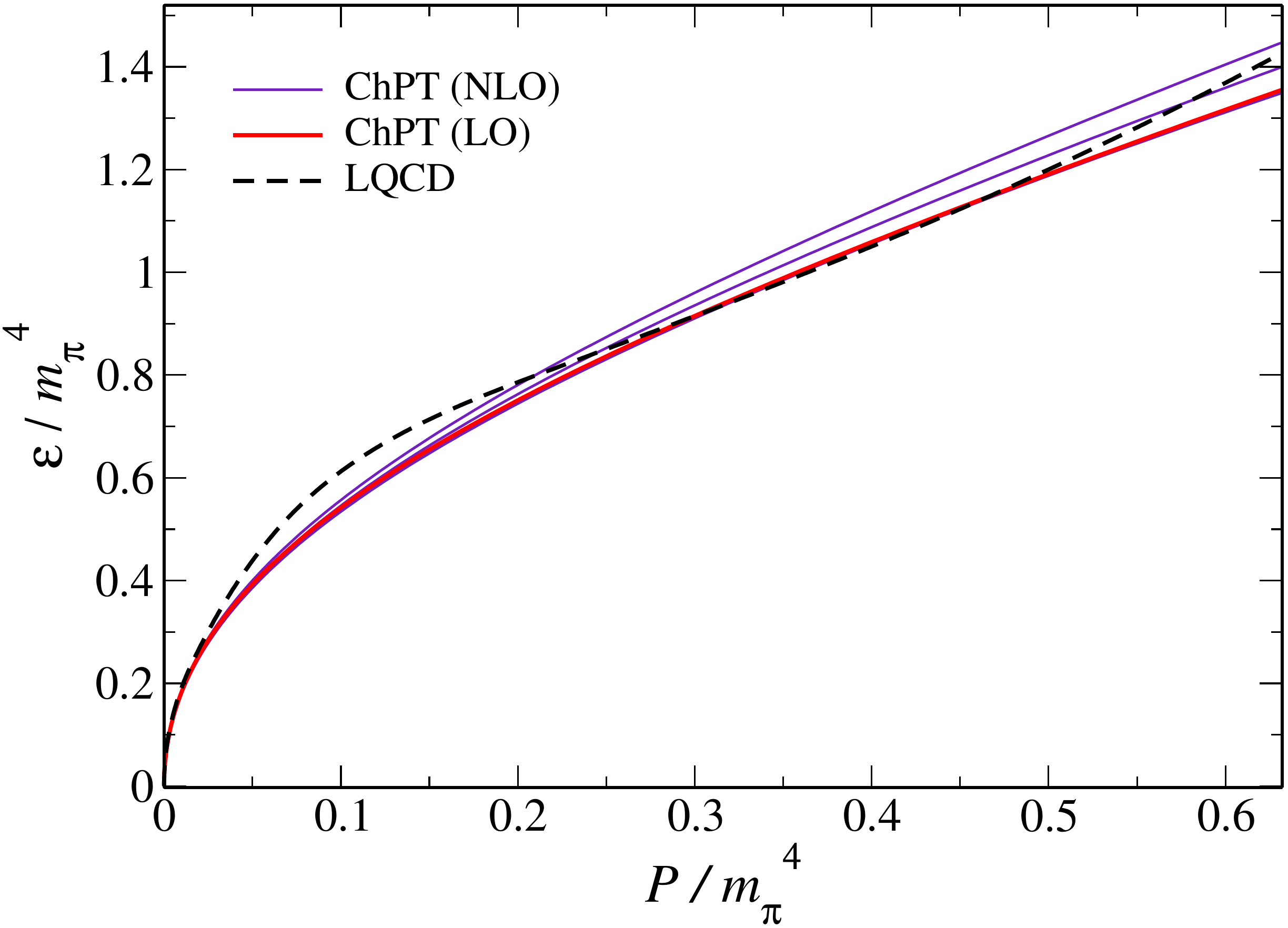}}
  \caption{Normalized equation of state. The LQCD results~\cite{Brandt:2018bwq} have been
  compared with the behavior of MSS and TRS within the NJL model (left panel) and with up to NLO results within $\chi$PT~\cite{Adhikari:pc} (right panel).
  The plots are specifically zoomed into the region of interest, up to the value of $\mu_I$ for which LQCD 
  data is available. The three lines for $\chi$PT-NLO depicts the uncertainty in the result due to the uncertainty in the low-energy constants~\cite{Adhikari:2019mdk,Adhikari:pc}. }
  \label{eos_comparison}
 \end{center}
\end{figure*}
Normalized EoS is presented in figure \ref{eos_comparison} where we can notice the reflection of 
the behavior of figures \ref{ni_comparison} and \ref{ed_comparison} regarding the comparability of 
NJL and LQCD results. As it can be seen, within the limit of their uncertainties NLO $\chi$PT results are in better agreement with the LQCD results 
for the region $P>0.2m_\pi^4$, whereas NJL (TRS and MSS) results are in better agreement in the lower region of $P<0.2m_\pi^4$. 

\begin{figure*}[!]
 \begin{center}
 \subfigure{\includegraphics[scale=0.33]{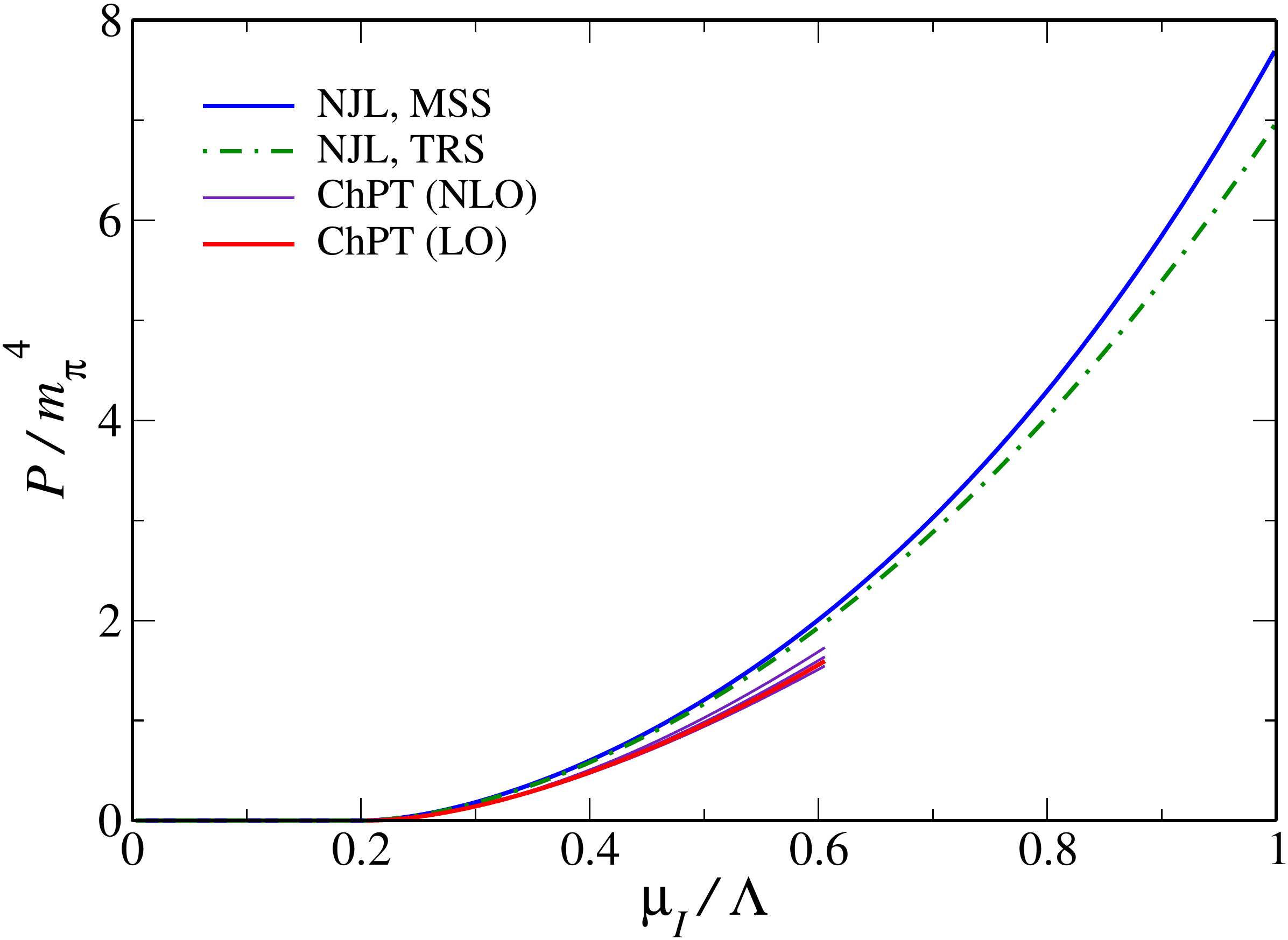} }
 \subfigure{\includegraphics[scale=0.33]{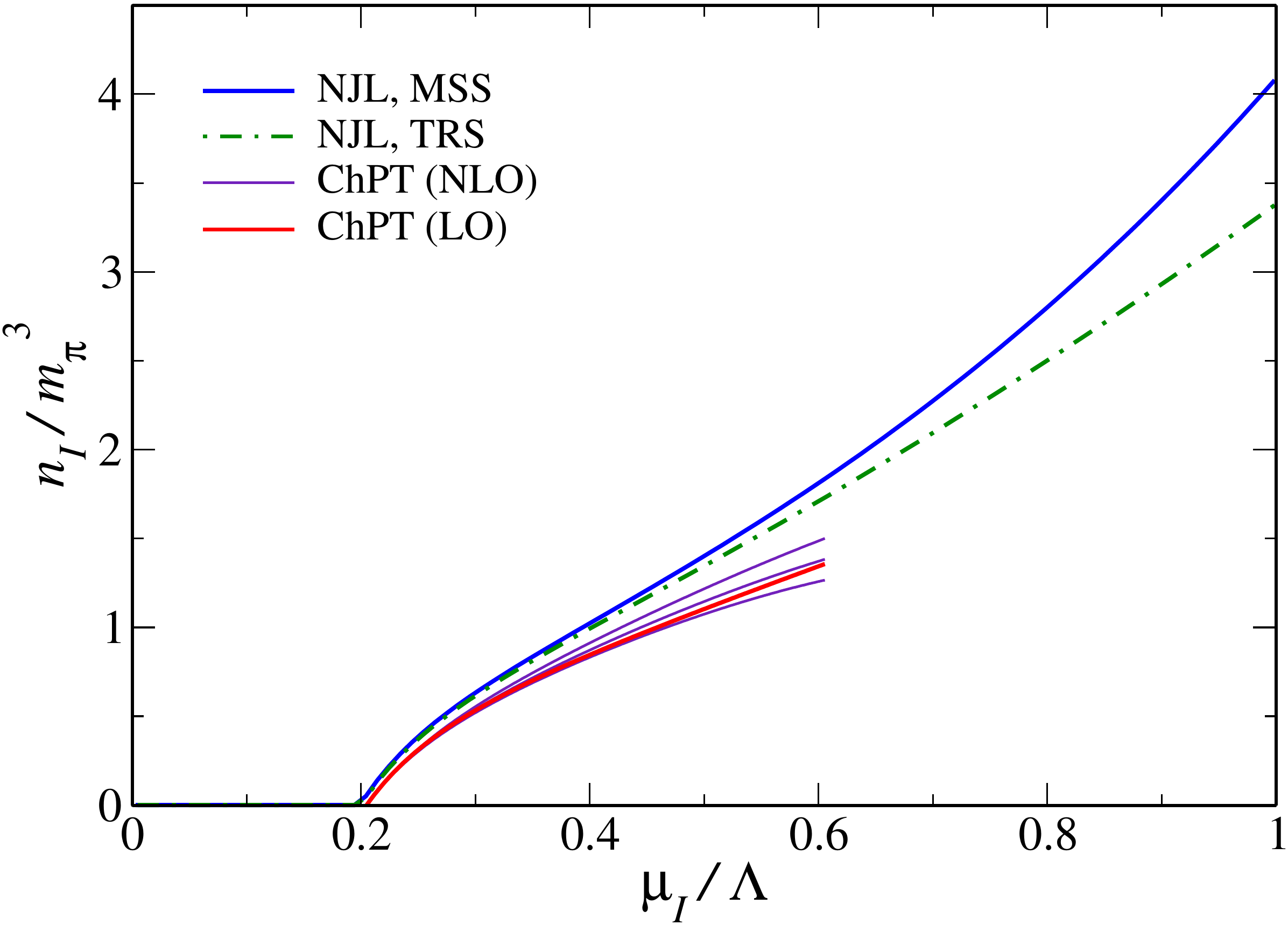} }
 \subfigure{\includegraphics[scale=0.33]{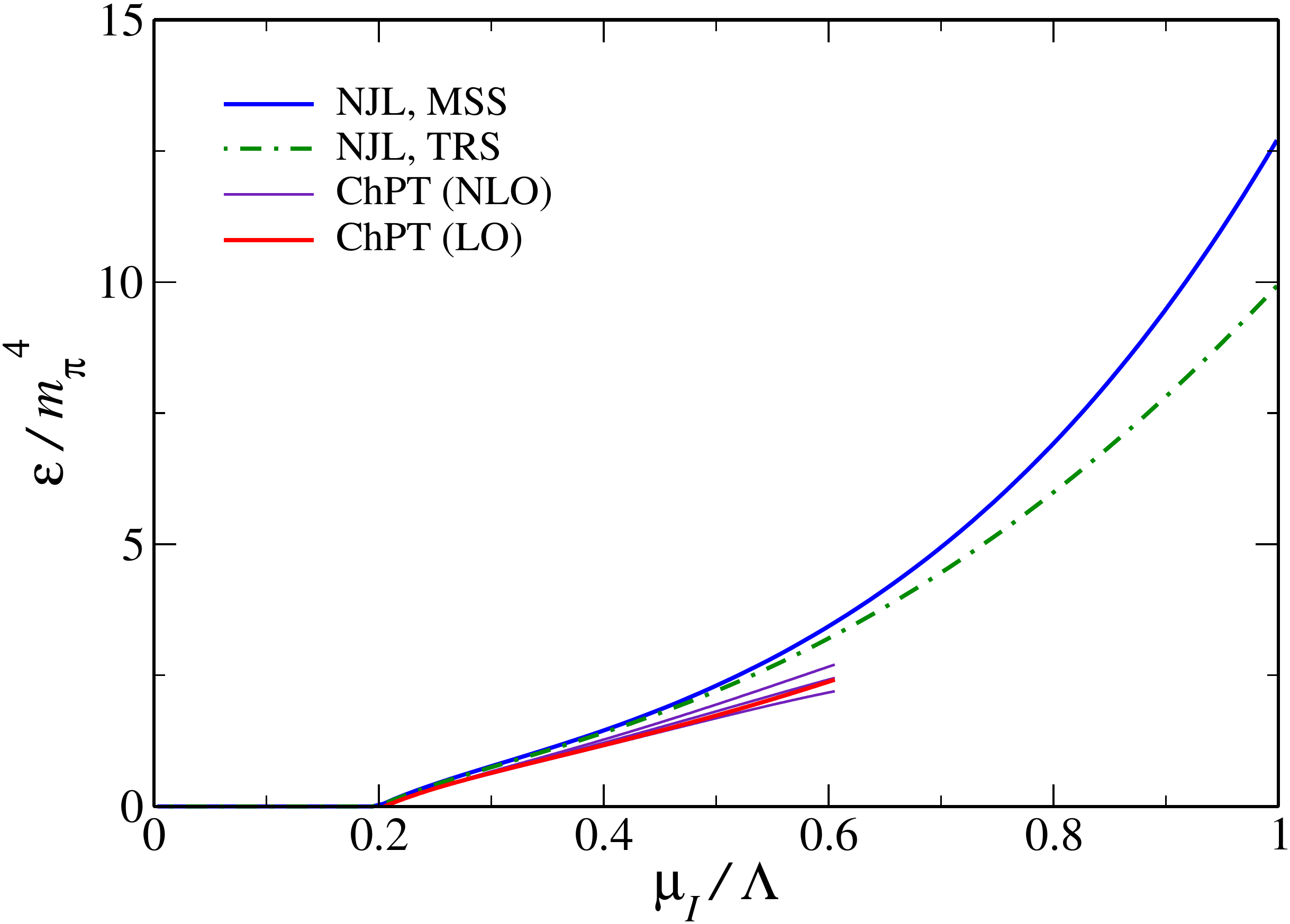} }
 \subfigure{\includegraphics[scale=0.33]{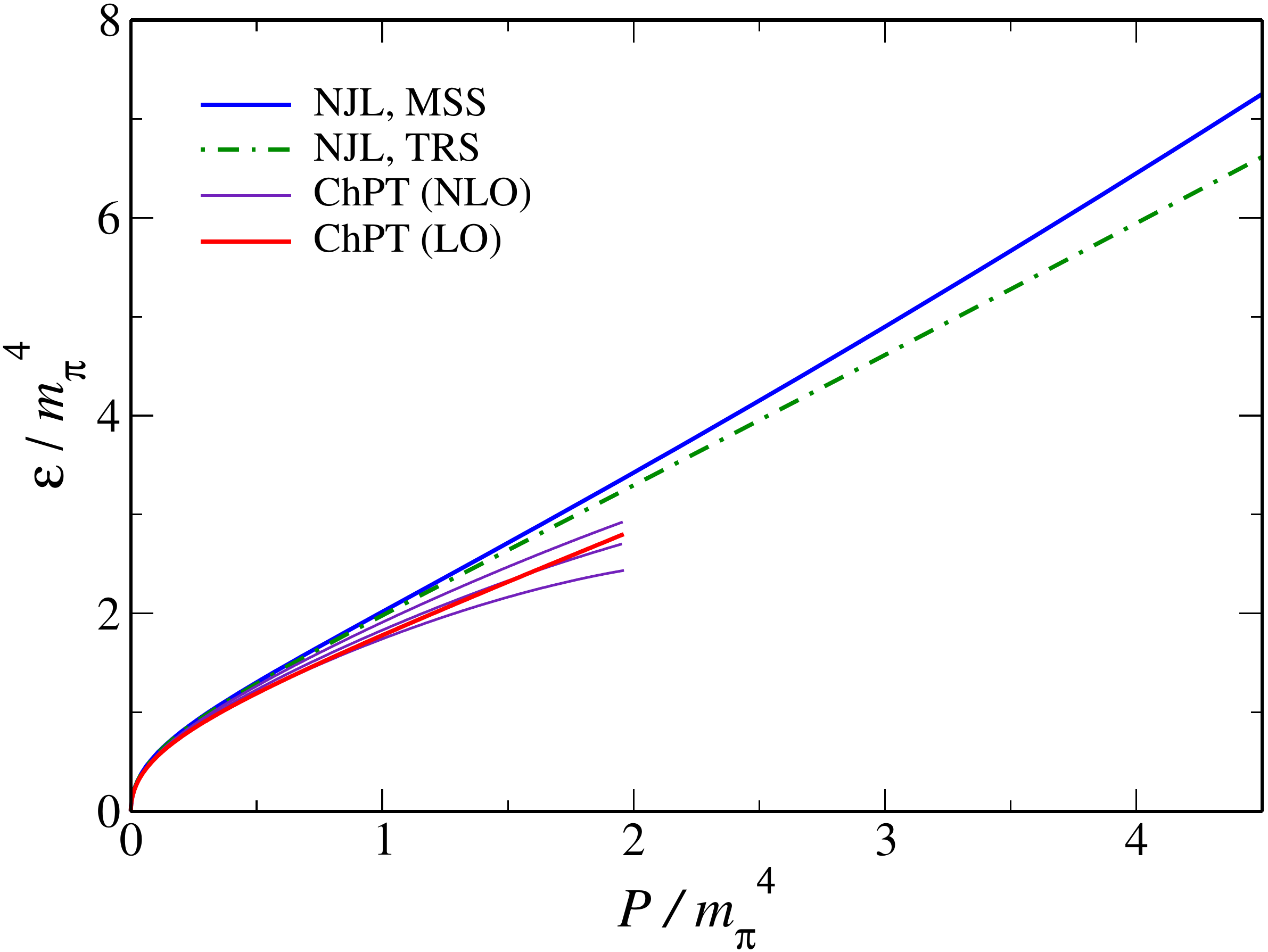} }
 \caption{Variations of the normalized pressure (upper left panel), isospin density (upper right panel), and energy density (lower left panel) are shown as a function of isospin chemical potential scaled with the 3D momentum cutoff ($\mu_I/\Lambda$) along with the normalized EoS (lower right panel). This plot shows the different behaviors of MSS and TRS within NJL model over the full spectrum of $\mu_I$ up to $\Lambda$. $\chi$PT results up to NLO have also been presented up to $\mu_I=0.6\Lambda$. The three lines for $\chi$PT-NLO depicts the uncertainty in the result due to the uncertainty in the low-energy constants~\cite{Adhikari:pc}.}
 \label{predictions}
 \end{center}
\end{figure*}

Finally in figure \ref{predictions} we consider the full spectrum of $\mu_I$, i.e. $<0\le\mu_I\le\Lambda$ to emphasize the effect of the medium separation at higher values of $\mu_I$ on the normalized thermodynamic
quantities $P_{\textrm{NJL}}$, $\langle n_I\rangle_{\textrm{NJL}}$ and $\varepsilon_{\textrm{NJL}}$ as well as the EoS. We interprete the parameter $\Lambda$ as the scale of the model, trusting in results restricted by $\Lambda$. In general we use this $\Lambda$ as an upper limit for the other relevant variables, e.g., temperature, external fields, chemical potentials etc, and the same idea was applied for $\mu_I$ in this work. Though it is true that for $\mu_I=\Lambda$ the regime of validity of our model ends, but we can see in Fig.~\ref{predictions} that the MSS results are different from TRS even for $\mu_I < \Lambda$. We have also plotted $\chi$PT results up to NLO in figure \ref{predictions} but only up to $\mu_I=0.6\Lambda$ ($\sim 3m_\pi$). This is to emphasize the fact that those results cannot be trusted beyond $\mu_I\sim 3m_\pi$ due to constraints on their validity~\cite{Adhikari:pc}.

\section{Conclusions}
\label{sec4}

In conclusion, we would like to emphasize on the fact that both the TRS and the MSS 
regularization schemes within the
NJL model show
promising results in the front of thermodynamic quantities describing systems similar to pion stars, being
largely in agreement with the LQCD results. Regions with higher values of $\mu_I$, where LQCD results are
not available, we have predicted the pressure, isospin density, energy density and EoS both within TRS and MSS,
highlighting the fact that MSS is more reliable in those regions
due to its unique way of
separating vacuum divergent 
effects from medium terms. In comparison with other effective theory results, i.e. $\chi$PT, our results within
the mean field NJL model show a better agreement with LQCD results
which prompts us to further investigate the
phase diagram for the region with finite $\mu_B$ and $\mu_I$ which is inaccessible by LQCD due to the sign problem.
Also as mentioned in section \ref{sec1}, the possibility of pion condensation in light of early universe dictates 
further exploration in the $T-\mu_I$ plane of the QCD phase diagram. Furthermore, $\chi$PT calculations for $SU(3)$ at finite isospin have also appeared very recently in~\cite{chptsu3}, which shows excellent agreement with lattice data for small values of $\mu_I$. Works in these directions within the NJL model are in progress.

\textit{Note added} - While finishing the updated version of our paper we learned that a partially overlapping study was done by Zhen-Yan Lu, Cheng-Jun Xia and Marco Ruggieri~\cite{Lu:2019diy}. 

\section{Acknowledgements}
This work was partially supported by Conselho Nacional de Desenvolvimento Cient\'{i}fico e Tecnol\'{o}gico(CNPq)
under grants 304758/2017-5 (R.L.S.F) and 6484/2016-1 (S.S.A) 
and as a part of the project INCT-FNA (Instituto Nacional de Ci\^encia e Tecnologia - F\'{\i}ısica Nuclear
e Aplica\c c\~oes) 464898/2014-5 (SSA), 
Coordenac\~{a}o de Aperfei\c{c}oamento de Pessoal de N\'{i}vel Superior (CAPES) (A.B) and Funda\c{c}\~ao de Amparo \`a Pesquisa do Estado de S\~ao Paulo (FAPESP)
under Grant No. 2017/26111-4 (D.C.D).

\end{document}